\documentclass[a4paper,11pt,dvips]{article}
\usepackage{dcolumn,graphicx,psfrag3,overcite,url}

\newcolumntype{d}{D{.}{.}{-1}}
\newcolumntype{L}{>{$}l<{$}}
\newcolumntype{R}{>{$}r<{$}}
\newcolumntype{C}{>{$}c<{$}}

\def\M#1{$\rm M_{#1}$}
\def\fsz{\footnotesize}
 
\begin{document}
 
\title{Structural relaxation in Morse clusters: Energy landscapes}
\author{Mark A.~Miller, Jonathan P.~K.~Doye and David J.~Wales}
\date{\it University Chemical Laboratory, Lensfield Road, \\Cambridge CB2 1EW, U.K.}
\maketitle

\begin{abstract}
We perform a comprehensive survey of the potential energy landscapes of
13-atom Morse clusters, and describe how they can be characterized
and visualized. Our aim is to detail how the global features of the
funnel-like surface change with the range of the potential, and to relate
these changes to the dynamics of structural relaxation. We find that the
landscape becomes rougher and less steep as the range of the potential
decreases, and that relaxation paths to the global minimum become more
complicated.
\end{abstract}

\section{Introduction}

Structural relaxation plays a key role in a diverse range of problems
in chemical physics, including protein folding, glass formation,
and the observation of ``magic number'' peaks in
the mass spectrometry of rare gas clusters. The dynamic evolution of such
systems is determined by the potential energy surface (PES) generated
by the interactions between their constituent particles. Quite
often one wants to find the structure and physical properties of
a (macro)molecule or cluster, by which it is usually meant the
properties of the global minimum on the PES, or, equivalently, the
properties at zero Kelvin.
However, the dynamics of a system at
temperatures or energies above which it can escape from the global
minimum depend on larger regions of the PES, the topology
and topography of which determine the precise behavior. When
considering the wider features of the PES in this way, it 
has become usual to refer to the PES as the ``potential energy landscape''.
\par
One can also consider the {\em free energy} landscape, a temperature-dependent
function which incorporates the entropy. For example, in protein
folding such a landscape can be defined either as a function of the protein
configuration by averaging the free energy over all solvent coordinates, or as
a function of distance from the folded state in terms of a similarity
parameter.\cite{Bryngelson95a}
\par
In recent years, much understanding has been gained in a number of
fields by relating structural and dynamical properties to the underlying PES.
For example, many years ago Levinthal pointed out the apparent
contradiction between the astronomical number of possible configurations
that a protein can adopt and the rapidity with which it finds
the biologically active structure when it folds.\cite{Levinthal69a,Levinthalnote}
The ``paradox'' is resolved by
realizing that efficient folding is only possible when the potential energy
landscape is dominated by a funnel, i.e.~consists largely of convergent kinetic
pathways leading down in energy towards the required
structure.\cite{Leopold92a} The precise features of a funnel may
vary, but the native state must be thermodynamically stable
at temperatures or energies where the dynamics are fast enough
for the system to be able to explore the landscape and find
it.\cite{Wolynes96a} The native state is destabilized if there are
structurally distinct states of low energy which can act as
kinetic traps.\cite{Doye96b} Hence, a pronounced global minimum encourages
efficient folding.\cite{Sali94b,Amara95a}
\par
The potential energy landscape also plays an important role in
determining the behavior of bulk liquids. Angell has proposed
a widely used scheme in which liquids are classified from
``strong'' to ``fragile''.\cite{Angell91a} A strong liquid
is characterized by a viscosity whose temperature dependence
follows an Arrhenius relationship ($\propto\exp[A/T]$). These
are often liquids with open network structures like water and
SiO$_2$, whereas fragile liquids tend to have more isotropic interactions.
Angell\cite{Angell95a} and Stillinger\cite{Stillinger95a} have
described the general features of the energy landscapes
that might be expected to characterize the two extremes. In a recent study,
Sastry et al.~have investigated the role of different regions of
the landscape in the process of glass formation in a model
fragile liquid.\cite{Sastry98a} They find that as the temperature of the liquid
is decreased, the system samples regions with higher barriers,
and on further cooling it samples deeper minima and
non-exponential relaxation sets in.
\par
Another way that an energy landscape can be classified is 
as ``sawtooth-like'' or ``staircase-like''
depending on the energy difference between minima relative to
the barriers which separate them.\cite{Kunz95a,Ball96a}
For example, the ``structure-seeking'' properties of the
(KCl)$_{32}$ cluster (i.e.~its ability to find a rock salt
structure even when cooled rapidly) can be attributed to
downhill barriers which are low compared
to the potential energy gradient towards
crystalline minima, as in a staircase.
\par
In order to characterize an energy landscape, it is necessary
to make a survey of its important features: minima, transition states
and pathways. Since the number of such features increases at least
exponentially with the number of particles in the
system\cite{Stillinger82a}, it is impractical and undesirable
to catalogue them all for large
systems. Consequently, existing studies have usually
concentrated on analyzing what is hoped to be a representative
sample of minima and transition states.\cite{Kunz95a,Doye96c} In this
study we examine in detail the landscape of the 13-atom Morse cluster
(\M{13}), which is large enough to possess a complex PES, but is small
enough for us to make a nearly exhaustive list of its minima and
transition states.
This model system is especially interesting because the energy
landscape is dominated by a funnel, and the potential contains one
parameter which allows us to adjust the complexity of the PES.
Previous studies\cite{Hoare76a,Braier90a}
have shown that potential energy surfaces are simpler
for short-ranged potentials,
and the effects of the range on the morphology of global minima of
atomic clusters\cite{Doye95d,Doye97b} and the stability of simple liquids\cite{Doye96d,Doye96c}
have already received attention. The range of the potential also
affects phase behavior: in a study of 7-atom Morse clusters, Mainz and Berry
found that liquid-like and solid-like phase coexistence is less distinct
when the range of attraction is longer.\cite{Mainz96a}
\par
In this paper, we concentrate on finding useful ways to characterize and visualize
a complex PES, and in the Summary we comment on how the range of
the potential is likely to affect the relaxation properties of the cluster.
We are currently using the data collected in this study to perform master
equation dynamics on the system to address relaxation in detail.

\section{Exploring the Landscape}

The Morse potential\cite{Morse29a} can be written in the form
\begin{equation}
V=\sum_{i<j}V_{ij};\hspace*{10mm}
V_{ij}=e^{\rho(1-r_{ij}/r_{\rm e})}
[e^{\rho(1-r_{ij}/r_{\rm e})}-2]\epsilon,
\end{equation}
where $r_{ij}$ is the distance between atoms $i$ and $j$.
$\epsilon$ and $r_{\rm e}$ are the dimer well depth and equilibrium bond length,
and simply scale the PES without affecting its topology.
They can conveniently be set to unity and used as the units of energy and distance.
$\rho$ is a dimensionless
parameter which determines the range of the inter-particle forces, with low values
corresponding to long range. Physically meaningful values range at least from
$\rho=3.15$ and $3.17$ for sodium and potassium\cite{Girifalco59a} to
$13.62$ for C$_{60}$ molecules.\cite{Girifalco92a} When $\rho=6$, the
Morse potential has the same curvature as the Lennard-Jones potential
at the minimum.
\par
The first step in characterizing the PES is to map out the local minima and
the network of transition states\cite{Murrell68a}
and pathways that connects them. The eigenvector-following
technique\cite{Pancir74a,Cerjan81a,Wales94a}
can efficiently locate transition states (first order saddles)
by maximizing the energy along a specified direction,
while simultaneously minimizing in all other directions.
The minima connected to a given transition state
are defined by the steepest descent paths commencing
parallel and antiparallel to the transition vector (the Hessian eigenvector
with negative eigenvalue) at the transition state. Although
eigenvector-following can also be used for these minimizations, the pathways
are not necessarily the same,\cite{Wales94a} and may even lead
to a different minimum. Since both
the pathways and the connectivity are of interest here, we use a
steepest descent technique for minimizations, employing analytic second
derivatives, following Page and McIver.\cite{Page88a}
\par
Our algorithm for exploring the PES is similar to that used by Tsai
and Jordan in a study of small Lennard-Jones and water clusters.\cite{Tsai93b}
Starting from a known minimum:
\begin{enumerate}
\item Search for a transition state along the eigenvector with the
lowest eigenvalue.\label{tssearch}
\item Deduce the path through this transition state and the minima
connected to it.
\item Repeat from step \ref{tssearch} beginning antiparallel to the
eigenvector, and then in both directions along eigenvectors with
successively higher eigenvalues until a specified number, $n_{\rm ev}$,
of directions have been searched uphill.
\item Repeat from step \ref{tssearch} until $n_{\rm ev}$ modes
of all known minima have been searched.
\end{enumerate}
By taking steps directly between minima, this method avoids wasting
time on intra-well dynamics. Other methods for exploring
energy landscapes, such as molecular dynamics, can become trapped
in local minima, especially at low temperature, where there is a wide
separation in time scale between inter-well and intra-well motion.
The chosen value of $n_{\rm ev}$ clearly affects the thoroughness of the
survey, although even if all $(3N-6)$ vibrational modes of an $N$-atom cluster
are searched, there is no guarantee of finding every minimum and transition
state. In practice, the required computer time and storage
demand that $n_{\rm ev}$ be reduced for large $\rho$, since the complexity
of the PES increases dramatically as the range of the potential
decreases. However, one finds that searches from low-lying minima are
more likely to converge in a reasonable number of iterations, so the
above algorithm was augmented with searches along further eigenvectors
of lower-energy minima. We are confident that the databases generated for
$\rho=4$ and 6 are nearly exhaustive, and
although those for higher values of $\rho$ are
necessarily less complete, this approach still allows us to map out the PES
fairly comprehensively.
\par
Details of the searches and the resulting databases
for $\rho=4$, 6, 10 and 14 are summarized in Table \ref{table:database}.
The dramatic rise in the number of minima and transition states
found as the range of the potential decreases is the first indication
of the increasing complexity of the PES.\cite{Hoare76a}
The remainder of this paper
investigates in more detail the nature of these changes and some useful
ways of characterizing the landscapes.

\section{Topological Mapping}

When trying to describe an energy ``landscape'', one has already been forced
to use terminology appropriate to a surface in three-dimensional space,
and pictorial representations are usually restricted even further to
two dimensions. Visualizing a $3N$-dimensional object directly in such
a way has obvious limitations, yet
it is appealing to have an idea of ``what the surface looks like''.
\par
One helpful way of doing this is to use topological mapping to construct
a disconnectivity graph, as applied to a polypeptide by
Becker and Karplus.\cite{Becker97a} The analysis begins by mapping
every point in configuration space onto the local minimum reached
by following the steepest descent path.\cite{Stillinger82a} Thus,
configuration space is represented by the discrete set of minima,
each of which has an associated ``well'' of points which map onto it.
Although this approach discards information about the volume of phase
space associated with each minimum, the density of minima can
provide a qualitative impression of the volumes associated with
the various regions of the landscape.
\par
At a given total energy, $E$, the minima can be grouped into disjoint sets,
called basins (``super basins'' in Becker and Karplus' nomenclature),
whose members are mutually accessible at that energy.
In other words, each pair of minima in a basin are connected directly or
through other minima by a path whose energy never exceeds $E$,
but would require more energy to reach a minimum in another basin.
At low energy there is just one basin---that containing the global minimum.
At successively higher energies, more basins come into play as new
minima are reached. At still higher energies, the basins coalesce as higher
barriers are overcome, until finally there is just one basin containing
all the minima (provided there are no infinite barriers).
\par
The disconnectivity graph is constructed by performing the basin
analysis at a series of energies, plotted on a vertical scale. At
each energy, a basin is represented by a node, with lines joining
nodes in one level to their daughter nodes in the level below. The
choice of the energy levels is important; too wide a spacing and
no topological information is left, whilst too close a spacing
produces a vertex for every transition state and hides the longer range
structure of the landscape. The horizontal position of the
nodes is arbitrary, and can be chosen for clarity.
In the resulting graph, all branches terminate at local minima, while
all minima connected directly or indirectly to a node are mutually
accessible at the corresponding energy.
\par
\begin{figure}
\begin{psfrags}
\psfrag{-33}{$-33$}
\psfrag{-34}{$-34$}
\psfrag{-35}{$-35$}
\psfrag{-36}{$-36$}
\psfrag{-37}{$-37$}
\psfrag{-38}{$-38$}
\psfrag{-39}{$-39$}
\psfrag{-40}{$-40$}
\psfrag{-41}{$-41$}
\psfrag{-42}{$-42$}
\psfrag{-43}{$-43$}
\psfrag{-44}{$-44$}
\psfrag{-45}{$-45$}
\psfrag{-46}{$-46$}
\psfrag{-47}{$-47$}
\psfrag{r=4}{$\rho=4$}
\psfrag{r=6}{$\rho=6$}
\centerline{\includegraphics[width=13cm]{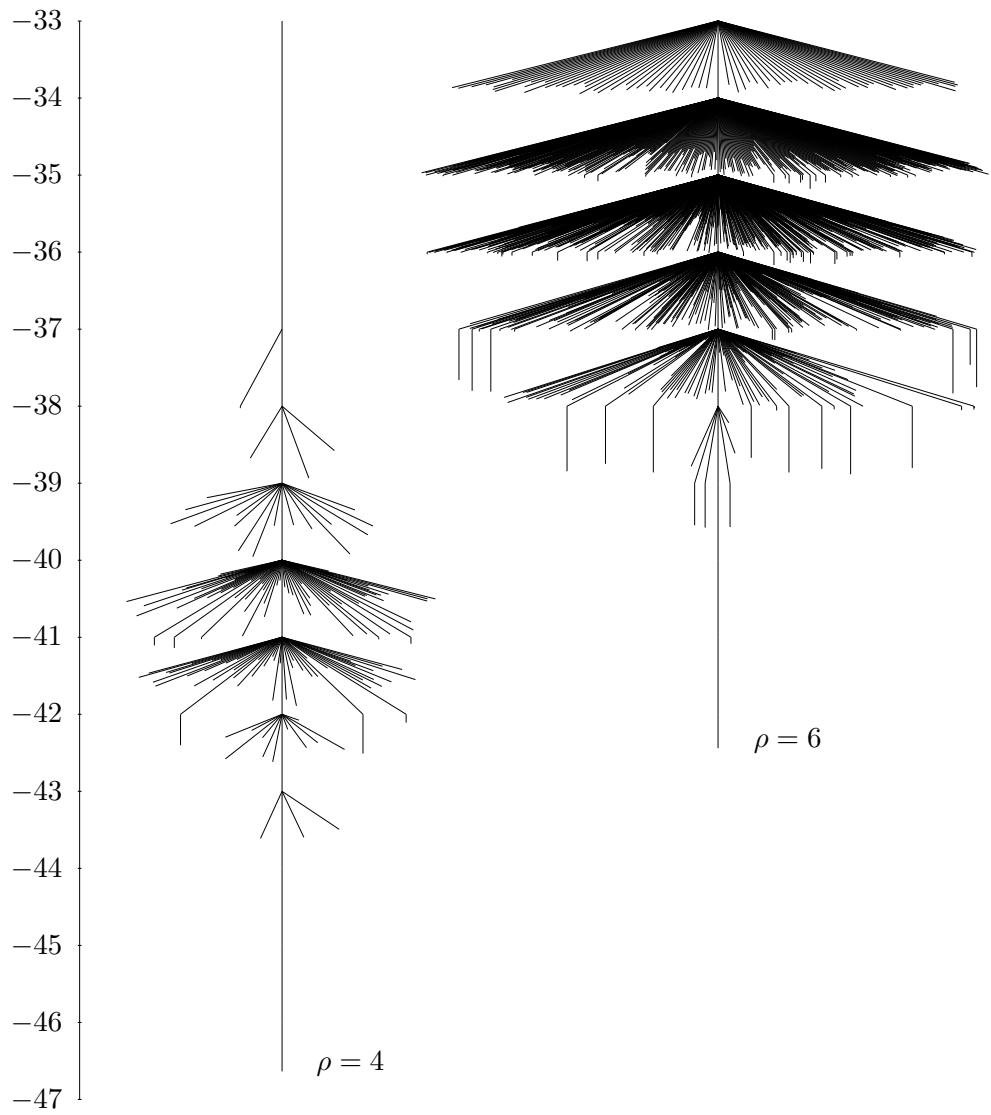}}
\end{psfrags}
\caption{Disconnectivity trees for \M{13} with $\rho=4$ and
$\rho=6$ plotted on the same energy scale (in units of the
pair well depth).}
\label{fig:trees}
\end{figure}

The disconnectivity graphs for \M{13} with $\rho=4$ and 6 are
plotted on the same scale in figure \ref{fig:trees}. We have chosen
a linear energy spacing of one well depth, which is an effective compromise
between the points raised above. Both trees are typical of a
funnel-like landscape: as the energy is lowered, minima are cut off a
few at a time with no secondary funnels, which would appear as
side branches. A large upward shift in the energy range of the minima
is apparent on increasing $\rho$ from 4 to 6, due to the increase in the
energetic penalty for strain and a decrease in the energetic contribution from 
next-nearest neighbors as the range of the potential decreases.\cite{Doye95d}
An increase in barrier
heights is also revealed by the somewhat longer branches at
$\rho=6$. Because of the large number of minima involved in the databases
for $\rho=10$ and 14, the disconnectivity graphs are too dense to
illustrate, but we shall see in the numerical analysis of the
next section how the trends develop.
\par
The concepts involved in the disconnectivity graph have much in common
with the ``energy lid'' description of Sibani et al.\cite{Sibani93a}
in which minima are grouped together if they are connected by paths never
exceeding a particular energy (the ``lid''). These authors plotted a tree
with a time axis, on which nodes represent the time when groups of
minima first come into equilibrium.
\par
The term ``basin'' has been used with a somewhat different meaning
by Berry and coworkers.\cite{Kunz95a,Berry95a} In this definition, a basin
consists of all minima connected to the basin bottom by a monotonic
sequence, i.e.~a sequence of connected minima with monotonically decreasing energy.
This definition contrasts with that of Becker and
Karplus,\cite{Becker97a} because it is independent of the energy, and
actually has a lot in common with the notion of a funnel.
Although the word
``funnel'' may conjure up a misleading image when the surface is rough
or shallow in slope, we will use it in this context to avoid confusion with
the previous definition of a basin as a set of mutually accessible minima
at a given energy. The funnel terminating at the global
minimum is denoted the primary funnel, whilst adjoining side funnels
are termed secondary. It should be noted that this definition
permits a minimum to belong to more than one funnel via different
transition states. The significance of dividing the landscape in this
way is that inter-funnel motion is likely to occur on a slower time scale
than inter-well flow,\cite{Kunz95a,Berry95a} so funnels constitute the next 
level in a hierarchy of landscape structure. Sufficiently deep or voluminous 
secondary wells can act as traps.\cite{Leopold92a,Doye96b}
A striking example is the cluster of 38 Lennard-Jones
atoms, whose truncated octahedral global minimum was only found
quite recently\cite{Doye95d,Pillardy95a} because of the much larger 
secondary funnel associated with a low-lying icosahedral structure.\cite{Wales98c}
\par
As the first line of Table \ref{table:landscape} shows, for $\rho=4$ the
landscape of \M{13}
is a perfect funnel: all minima lie on monotonic sequences terminating at
the global minimum. At higher values of $\rho$ a small fraction of
minima lie outside the primary funnel, and although they technically
constitute secondary funnels, they represent a very small proportion
of the phase space. We will now see how the characteristics of the
primary funnel evolve as the range of the potential is decreased.

\section{Properties of the Landscape}

The remainder of Table \ref{table:landscape} lists some global properties
of the landscape at four values of $\rho$. Some of the trends are
straightforward to understand.
For example, defining $\bar\nu_i$ as the geometric mean of the normal mode
frequencies at minimum $i$, the average of this quantity over the database
of minima, $\langle\bar\nu\rangle_{\rm m}$, rises monotonically with $\rho$ because
of the increasing stiffness of shorter-ranged potentials.
The average of the transition state imaginary frequency,
$\nu^{\rm im}_i$, increases less rapidly in magnitude, and
levels off at high $\rho$, indicating that the transition regions are flatter
relative to the well bottoms than at low $\rho$.
\par
\begin{figure}
\begin{psfrags}
\psfrag{energy}{\fsz ${\rm energy}/\epsilon$}
\psfrag{rho}{\fsz $\rho$}
\psfrag{-36}{\fsz $-36$}
\psfrag{-38}{\fsz $-38$}
\psfrag{-40}{\fsz $-40$}
\psfrag{-42}{\fsz $-42$}
\psfrag{-44}{\fsz $-44$}
\psfrag{-46}{\fsz $-46$}
\psfrag{-48}{\fsz $-48$}
\psfrag{3}{\fsz $3$}
\psfrag{5}{\fsz $5$}
\psfrag{7}{\fsz $7$}
\psfrag{9}{\fsz $9$}
\psfrag{11}{\fsz $11$}
\psfrag{13}{\fsz $13$}
\psfrag{15}{\fsz $15$}
\psfrag{Ih}{\fsz $I_h$}
\psfrag{D5h}{\fsz $D_{5h}$}
\psfrag{Cs}{\fsz $C_s$}
\centerline{\includegraphics[width=8.5cm]{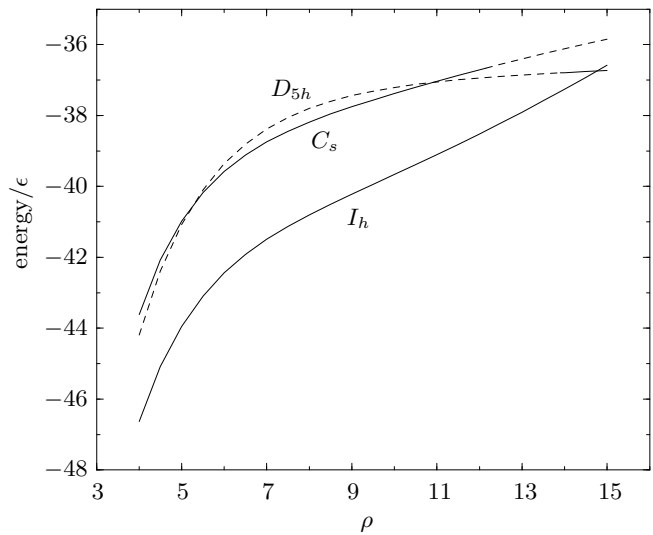}}
\end{psfrags}
\caption{Correlation diagram for some low-lying structures (see
figure \ref{fig:clusters}): the icosahedron
($I_h$), the decahedron ($D_{5h}$) and the lowest-energy defective
icosahedron ($C_s$). Dashed lines indicate regions where the structure
is not a minimum: $D_{5h}$ becomes a transition state and $C_s$ becomes a
second order saddle.}
\label{figure:correlation}
\end{figure}
\begin{figure}[t]
\begin{psfrags}
\psfrag{(a)}{\fsz (a)}
\psfrag{(b)}{\fsz (b)}
\psfrag{(c)}{\fsz (c)}
\centerline{\includegraphics[width=8.5cm]{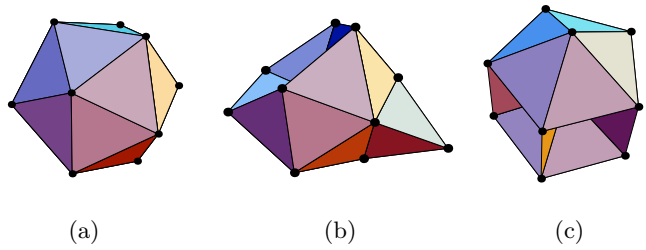}}
\end{psfrags}
\caption{Structures discussed in the text: (a) the icosahedron ($I_h$),
(b) the lowest-energy defective icosahedron ($C_s$), and (c) the
decahedron ($D_{5h}$).}
\label{fig:clusters}
\end{figure}

The increasing energy, $E_{\rm gm}$,
of the global minimum was noted in the previous section, and the table shows that
this increase is accompanied by a decreasing gap $\Delta E_{\rm gap}$ to the second lowest
minimum. The striking drop in $\Delta E_{\rm gap}$
when $\rho$ reaches 14 is due to a change in
morphology of the second lowest structure, as illustrated in figure
\ref{figure:correlation}. To see why this happens, it is helpful to decompose
the potential energy into the following contributions:\cite{Doye95b}
\begin{equation}
V=-n_{\rm nn} + E_{\rm strain} + E_{\rm nnn},
\label{eqn:decompose}
\end{equation}
where $n_{\rm nn}$ is the number of nearest neighbor interactions, i.e.
the number of pairs lying closer than a value $r_0$ (taken
here to be $1.15r_{\rm e}$), and the strain
energy and non-nearest neighbor contributions are defined by
\begin{equation}
E_{\rm strain} = \sum_{\stackrel{\scriptstyle i<j}{r_{ij}<r_0}}[V_{ij}+1],
\end{equation}
\begin{equation}
E_{\rm nnn} = \sum_{\stackrel{\scriptstyle i<j}{r_{ij}\ge r_0}}V_{ij}.
\end{equation}
$n_{\rm nn}$ and $E_{\rm strain}$ are more sensitive properties of the structure than
$E_{\rm nnn}$, and so the lowest energy cluster is determined by a balance
between maximizing $n_{\rm nn}$ and minimizing $E_{\rm strain}$.
The icosahedron [figure \ref{fig:clusters}(a)] is the global
minimum for all four values of $\rho$ considered here because it has 
the largest number of nearest neighbors ($n_{\rm nn}=42$). 
However, the large value of $n_{\rm nn}$ is at the expense of considerable strain.
As $E_{\rm strain}$ is the energetic penalty for nearest-neighbor distances deviating
from $r_{\rm e}$, it increases rapidly for strained structures as the pair-potential well
narrows at larger $\rho$. $E_{\rm nnn}$ is also sensitive to $\rho$; it 
decreases as the range of the potential decreases.
\par
The upward trends in figure \ref{figure:correlation} are caused by
the changes in $E_{\rm strain}$ and $E_{\rm nnn}$. For
$\rho<13.90$ the second lowest minimum is a defective icosahedron in which
one vertex has been removed and one face is capped [figure \ref{fig:clusters}(b)].
The removal of a
vertex allows the strain in the icosahedron to relax, and so the energy
rises less steeply than for the icosahedron and $\Delta E_{\rm gap}$
falls. However, decahedral clusters are intrinsically less strained
than icosahedral ones, and at $\rho=13.90$ the decahedron
[figure \ref{fig:clusters}(c)], which for
lower $\rho$ is a transition state, becomes the second lowest minimum.
In fact, for $\rho>14.77$ the decahedron is the global minimum, although
this value of $\rho$ may be too large to be observed in chemical systems.
The change in the order of the stationary point arises from a delicate
balance between $E_{\rm strain}$ and $E_{\rm nnn}$. The vibrational mode
of the decahedron
with the lowest Hessian eigenvalue is a twist about the $C_5$ axis.
This motion strains the structure, but brings non-nearest neighbors closer.
At high $\rho$, the increased strain wins, causing the energy to rise and
giving a minimum, whereas for longer-ranged interactions the non-nearest
neighbors lower the energy, giving a saddle.
\par
\begin{figure}
\begin{psfrags}
\psfrag{density of minima}{\fsz density of minima}
\psfrag{energy / e}{\fsz energy / $\epsilon$}
\psfrag{r=4}{\fsz $\rho=4$}
\psfrag{r=6}{\fsz $\rho=6$}
\psfrag{r=10}{\fsz $\rho=10$}
\psfrag{r=14}{\fsz $\rho=14$}
\psfrag{0}{\fsz 0}
\psfrag{5}{\fsz 5}
\psfrag{10}{\fsz 10}
\psfrag{15}{\fsz 15}
\psfrag{20}{\fsz 20}
\psfrag{40}{\fsz 40}
\psfrag{60}{\fsz 60}
\psfrag{80}{\fsz 80}
\psfrag{100}{\fsz 100}
\psfrag{500}{\fsz 500}
\psfrag{1000}{\fsz 1000}
\psfrag{1500}{\fsz 1500}
\psfrag{2000}{\fsz 2000}
\psfrag{3000}{\fsz 3000}
\psfrag{4000}{\fsz 4000}
\psfrag{-50}{\fsz $-50$}
\psfrag{-45}{\fsz $-45$}
\psfrag{-40}{\fsz $-40$}
\psfrag{-35}{\fsz $-35$}
\psfrag{-30}{\fsz $-30$}
\centerline{\includegraphics[width=18.0cm]{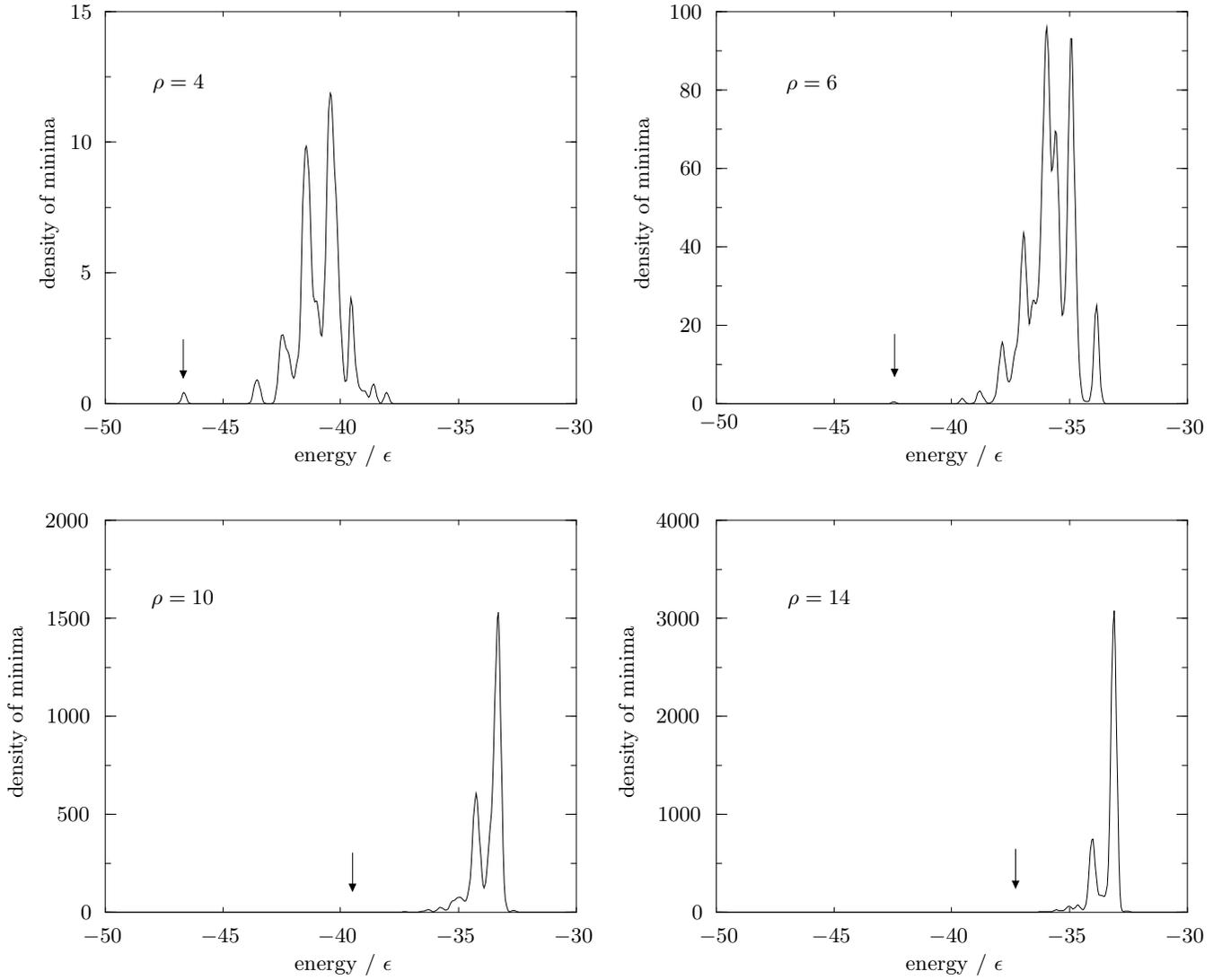}}
\end{psfrags}
\caption{Energy distribution of the minima for four values of the
range parameter $\rho$. In each case, the energy of the global minimum is
indicated by an arrow.}
\label{figure:mindist}
\end{figure}

The decreasing $\Delta E_{\rm gap}$ indicates a local flattening
of the PES at higher $\rho$. This effect extends beyond the vicinity of
the global minimum to the whole landscape, as
can be seen from the energy distributions of minima shown in figure
\ref{figure:mindist}. As $\rho$ increases, the energy distribution
shifts upwards and becomes narrower, and for $\rho=10$ and 14
it develops two sharp peaks at $-33$ and $-34$.
At high values of $\rho$,
$E_{\rm nnn}$ becomes small, and the energetic penalty for strain
is large. Decomposition of the energy according to equation
\ref{eqn:decompose} reveals that the peaks in the distributions
correspond to low-strain structures with 33 and 34 nearest neighbors.
Low strain can arise from two structural motifs: close packing or
polytetrahedral packing (without pentagonal bipyramids). It is not easy to
classify such a small cluster according to these schemes, but it is worth
noting that the radial distribution function, taken over all the
minima, develops a $\sqrt2$ signature as $\rho$ increases, which
is characteristic of close packing.\cite{Ashcroft76a}
\par
A more quantitative measure of the slope of the PES is provided by
the energy difference between pairs of
connected minima, $\Delta E^{\rm con}_i$
(where $i$ labels the connecting transition state, or, equivalently
the pathway). As Table \ref{table:landscape} shows, the average of
this quantity over the pathways
drops off quickly as $\rho$ increases from 4. $\Delta E^{\rm con}_i$ is
the difference between the uphill and downhill barriers $b^{\rm up}_i$
and $b^{\rm down}_i$ defined by transition state $i$ and the two minima
it connects. Although the average over the pathways of the uphill barrier,
$\langle b^{\rm up}\rangle_{\rm p}$,
decreases as the range of the potential decreases,
$\langle b^{\rm down}\rangle_{\rm p}$
increases, i.e.~the barriers that must be overcome for structural
relaxation towards the global minimum are larger; the flattening
of the funnel is accompanied by roughening.
\par
Given the dramatic increase in the number of stationary points as
the range of the potential decreases, and that the volume of
accessible phase space will be reduced as the long range attraction
is squeezed out, we must expect some change in the nature of the
individual pathways between minima and their organization on
the landscape.
Defining $D_i$ as the separation in configuration space of the
two minima connected by transition state $i$,
Table \ref{table:landscape} shows, as we might
expect, that connected minima are on average closer
when the potential is short-ranged. This effect is accompanied by
a decrease in the average of the integrated path length, $S_i$.
It is interesting to see how the individual pathways are organized
into routes to the global minimum. We have calculated the shortest
path from each minimum to the global minimum, as measured by the
total integrated path length $S^{\rm gm}_i$ (the path with fewest
{\em steps} between minima is generally longer). The average of
$S^{\rm gm}_i$ is fairly insensitive to $\rho$, whilst the
average of the number of steps along the corresponding
pathways, $n^{\rm gm}_i$, increases. Thus,
on average, the path for relaxation to the global minimum does not
increase significantly in length, but becomes more rugged as more
transition states must be crossed. Whereas every minimum at
$\rho=4$ can reach the global minimum in either 1 or 2 steps,
as many as 5 may be required at $\rho=14$. Table \ref{table:steps} shows
how the minima are distributed over $n^{\rm gm}_i$, giving some
insight into the connectivity of the landscape. The number of minima
with $n^{\rm gm}_i=1$ tells us how many transition states are connected
directly to the global minimum. The values are remarkably high, especially
as permutational isomers are not included. Interestingly,
the number of minima does not increase continuously as the sequences
branch out from the global minimum (as one might expect in a funnel),
but tails off quite gently.
\par
An intuitive explanation for the constancy of $\langle S^{\rm gm}\rangle$
and the increase in $\langle n^{\rm gm}\rangle$ might be that paths are split
into a larger number of sub-rearrangements. The number of atoms
contributing to rearrangement $i$ can be measured by the cooperativity
index $\tilde N_i=N/\gamma_i$, where $\gamma_i$ is the moment ratio
of displacement, which is defined by\cite{Stillinger83a}
\begin{equation}
\gamma_i={N\sum_\alpha^N|{\bf r}_\alpha(s)-{\bf r}_\alpha(t)|^4\over
\left(\sum_\alpha^N|{\bf r}_\alpha(s)-{\bf r}_\alpha(t)|^2\right)^2},
\end{equation}
where ${\bf r}_\alpha$ is the Cartesian position vector of atom $\alpha$,
and $s$ and $t$ denote
the final and initial configurations in rearrangement pathway $i$.
Table \ref{table:landscape} shows that the average
value of $\tilde N_i$ is almost independent of $\rho$. In fact the
distribution of $\tilde N_i$ (from 1 to $N$) is remarkably similar
for all four databases. This result contrasts with statistics previously
obtained for
the larger clusters LJ$_{55}$ and $({\rm C}_{60})_{55}$, which
showed that cooperative (high $\tilde N_i$) rearrangements
are less likely for $({\rm C}_{60})_{55}$, where the
range of the potential corresponds to $\rho\approx 14$.\cite{Wales94a}
It is possible that a 13-atom
cluster is too small to support localized sub-rearrangements
in this way.

\section{Summary}

We have performed a comprehensive survey of the potential energy
landscapes of the 13-atom Morse cluster for four values of the
range parameter using systematic eigenvector-following
searches. The landscapes were then characterized in detail
using disconnectivity graphs, funnel analysis, and
a selection of parameters that provide insight into the topology and
topography. We have described and rationalized the changes in the
landscape as the range of the potential is varied over a physically
meaningful range.
\par
\begin{figure}
\begin{psfrags}
\psfrag{energy / e}{\fsz ${\rm energy}/\epsilon$}
\psfrag{S / s}{\fsz $S/\sigma$}
\psfrag{-30}{\fsz $-30$}
\psfrag{-35}{\fsz $-35$}
\psfrag{-40}{\fsz $-40$}
\psfrag{-45}{\fsz $-45$}
\psfrag{-50}{\fsz $-50$}
\psfrag{r=4}{\fsz $\rho=4$}
\psfrag{r=6}{\fsz $\rho=6$}
\psfrag{r=14}{\fsz $\rho=14$}
\psfrag{-10}{\fsz\ 10}
\psfrag{-5}{\fsz\ 5}
\psfrag{0}{\fsz 0}
\psfrag{5}{\fsz 5}
\psfrag{10}{\fsz 10}
\centerline{\includegraphics[width=8.5cm]{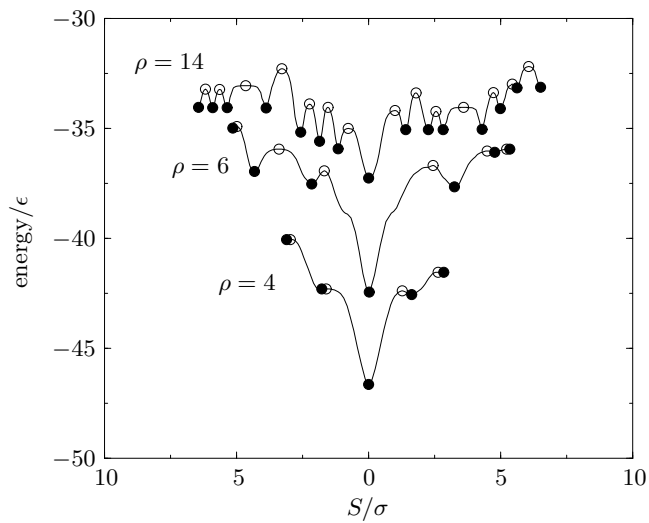}}
\end{psfrags}
\caption{Example monotonic sequences leading to the global minimum
for three values of the range parameter $\rho$. $S$ is the integrated
distance along the reaction
path from the global minimum. Minima are indicated by filled circles, and transition
states by open circles. The plots demonstrate a number of features discussed in
the text: the general increase in energy of the minima, the decreasing gap to the
global minimum, the increasing barrier heights, the shorter rearrangements,
and the decreasing gradient towards the global minimum as the range of the
potential decreases.}
\label{figure:monotonic}
\end{figure}

The trends displayed in Table
\ref{table:landscape} and the above discussion are underlined
by the plots of representative monotonic sequences in figure
\ref{figure:monotonic}. The overall classification of the potential
energy landscape is that of a funnel, but one which becomes
flatter and rougher as the range of the potential
decreases. This change is accompanied by a general increase in complexity
of the surface in terms of the number of minima and transition
states and in the number of steps required to reach one minimum from another.
\par
Previous studies of model potential landscapes\cite{Doye96b} have shown
that relaxation from high-energy configurations to the global minimum 
is most efficient when the PES has a large potential energy
gradient towards the
global minimum with low downhill barriers, and lacks secondary funnels
which act as kinetic traps. On this basis we would expect \M{13} to
relax most easily when the the range of the potential is long, in
spite of the fact that the frequency of intra-well vibrational oscillations
decreases as the potential becomes less ``stiff'' at fixed values of
$\epsilon$ and $r_{\rm e}$ (see Table \ref{table:landscape}). Low values
of the range parameter $\rho$ are therefore likely to produce
``structure-seekers'' whereas high values will tend to produce
``glass-formers'', reflecting a continuous change from a staircase-like
to a sawtooth-like landscape.
\par
Equipped with an understanding of the potential energy landscape and its
dependence on the range of the potential, we have applied
the master equation approach to investigate the dynamics of
structural relaxation in \M{13}.
This work enables us to probe in detail the flow of probability
between individual minima in an ensemble of clusters as they
approach the equilibrium distribution, and the results will be
described in a separate publication.

\vspace*{7mm}
\noindent
{\Large\bf Acknowledgments}
\vspace*{4mm}

M.A.M.~and D.J.W.~gratefully acknowledge financial support from
the Engineering and Physical Sciences Research Council, and the
Royal Society, respectively. 
M.A.M.~would also like to thank Gonville
\& Caius College, Cambridge for a grant towards attending the workshop
on energy landscapes at the Telluride Summer Research Center in July 1997.
 
\def\aciee{Angew. Chem. Int. Ed. Engl.}
\def\ac{Acta. Crystallogr.}
\def\acp{Adv. Chem. Phys.}
\def\acr{Acc. Chem. Res.}
\def\ajp{Am. J. Phys.}
\def\ap{Ann. Physik}
\def\arpc{Ann. Rev. Phys. Chem.}
\def\bmk{Biometrika}
\def\cccc{Coll. Czech. Chem. Comm.}
\def\cp{Chem. Phys.}
\def\cpc{Comp. Phys. Comm.}
\def\cpl{Chem. Phys. Lett.}
\def\crev{Chem. Rev.}
\def\el{Europhys. Lett.}
\def\fd{Faraday Discuss.}
\def\ic{Inorg. Chem.}
\def\ijmpc{Int. J. Mod. Phys. C}
\def\ijqc{Int. J. Quant. Chem.}
\def\jcis{J. Colloid Interface Sci.}
\def\jcsft{J. Chem. Soc., Faraday Trans.}
\def\jacs{J. Am. Chem. Soc.}
\def\jas{J. Atmos. Sci.}
\def\jcc{J. Comp. Chem.}
\def\jchp{J. Chim. Phys.}
\def\jcp{J. Chem. Phys.}
\def\jce{J. Chem. Ed.}
\def\jcscc{J. Chem. Soc., Chem. Commun.}
\def\jetp{J. Exp. Theor. Phys. (Russia)}
\def\jmb{J. Mol. Biol.}
\def\jmsp{J. Mol. Spec.}
\def\jmst{J. Mol. Struct.}
\def\jncs{J. Non-Cryst. Solids}
\def\jpa{J. Phys. A}
\def\jpb{J. Phys. B}
\def\jpc{J. Phys. Chem.}
\def\jpca{J. Phys. Chem. A}
\def\jpcb{J. Phys. Chem. B}
\def\jpcm{J. Phys. Condensed Matter.}
\def\jpcs{J. Phys. Chem. Solids.}
\def\jpsj{J. Phys. Soc. Jpn.}
\def\mg{Math. Gazette}
\def\mp{Mol. Phys.}
\def\nat{Nature}
\def\pac{Pure. Appl. Chem.}
\def\phys{Physics}
\def\pma{Philos. Mag. A}
\def\pmb{Philos. Mag. B}
\def\pnasu{Proc. Natl. Acad. Sci. USA}
\def\pr{Phys. Rev.}
\def\prep{Phys. Reports}
\def\pra{Phys. Rev. A}
\def\prb{Phys. Rev. B}
\def\prc{Phys. Rev. C}
\def\prd{Phys. Rev. D}
\def\pre{Phys. Rev. E}
\def\prl{Phys. Rev. Lett.}
\def\prsa{Proc. R. Soc. A}
\def\psfg{Proteins: Struct., Func. and Gen.}
\def\ptps{Prog. Theor. Phys. Supp.}
\def\sci{Science}
\def\sa{Sci. Amer.}
\def\ss{Surf. Sci.}
\def\tca{Theor. Chim. Acta}
\def\zpb{Z. Phys. B.}
\def\zpc{Z. Phys. Chem.}
\def\zpd{Z. Phys. D}

\bibliographystyle{thesis}

\newpage

%
%
\begin{table}[h]
\caption{Details of the databases for \M{13} at four values of the range
parameter $\rho$.
$n_{\rm ev}$ is the minimum number of eigenvectors of each minimum searched for a
transition state, and $n_{\rm s}$ is the average number of searches from each
minimum. $n_{\rm min}$ and $n_{\rm ts}$ are the numbers of minima
and transition states found.}
\begin{center}
\begin{tabular}{Ldddd}
\hline\hline
\rho           &   4  & 6      & 10        & 14      \\
\hline
n_{\rm ev}     & 15   & 6      & 3         & 2       \\
n_{\rm s}      & 31.3 & 13.0   & 7.0       & 7.5     \\
n_{\rm min}    & 159  & 1\,439 & 9\,306    & 12\,760 \\
n_{\rm ts}     & 685  & 8\,376 & 37\,499   & 54\,439 \\
\hline\hline
\end{tabular}
\label{table:database}
\end{center}
\end{table}

%
%
\begin{table}
\caption{Some properties of the potential energy landscape of \M{13} at four
values of the range parameter $\rho$. All dimensioned quantities are
tabulated in reduced units. $n_{\rm min}$ is the number of minima, of which
$n_{\rm pf}$ lie in the primary funnel. $E_{\rm gm}$ is the energy
of the global minimum, with the next-lowest energy structure lying
$\Delta E_{\rm gap}$ higher. $\bar\nu_i$ is the geometric mean normal
mode frequency at minimum $i$ and $\nu^{\rm im}_i$ is the imaginary frequency
at transition state $i$. $b^{\rm up}_i$ is the larger (uphill) barrier height between
the two minima connected by transition state $i$,
$b^{\rm down}_i$ is the smaller (downhill) barrier, and
$\Delta E^{\rm con}_i$ is the energy difference between the minima,
so that $b^{\rm up}_i=b^{\rm down}_i + \Delta E^{\rm con}_i$. $S_i$ is the integrated
path length between the two minima connected by transition state $i$,
$D_i$ is their separation in
configuration space, and $\tilde N_i$ is the cooperativity index of the
rearrangement (defined in the text). $n^{\rm gm}_i$ is the
smallest number of steps from minimum $i$ to the global minimum, and
$S^{\rm gm}_i$ is the integrated length of this path.
$\langle\cdots\rangle_{\rm m}$, $\langle\cdots\rangle_{\rm ts}$
and $\langle\cdots\rangle_{\rm p}$ indicate averages where the index
runs over minima,
transition states, and non-degenerate pathways (i.e.~pathways not merely
connecting permutational isomers) respectively.}
\begin{center}
\begin{tabular}{Ldddd}
\hline\hline
\rho                                     & 4       & 6       & 10      & 14      \\
\hline
n_{\rm min} - n_{\rm pf}                 & 0       & 1       & 219     & 442     \\
E_{\rm gm}                               & -46.635 & -42.440 & -39.663 & -37.259 \\
\Delta E_{\rm gap}                       & 3.024   & 2.864   & 2.245   & 0.468   \\
\langle\bar\nu\rangle_{\rm m}            & 1.187   & 1.625   & 2.615   & 3.660   \\
\langle|\nu^{\rm im}|\rangle_{\rm ts}    & 0.396   & 0.473   & 0.637   & 0.628   \\
\langle b^{\rm up}\rangle_{\rm p}        & 3.666   & 2.070   & 1.470   & 1.536   \\
\langle b^{\rm down}\rangle_{\rm p}      & 0.461   & 0.543   & 0.583   & 0.784   \\
\langle\Delta E^{\rm con}\rangle_{\rm p} & 3.205   & 1.526   & 0.887   & 0.752   \\
\langle S\rangle_{\rm p}                 & 2.457   & 1.735   & 1.030   & 0.971   \\
\langle D\rangle_{\rm p}                 & 1.462   & 1.163   & 0.840   & 0.817   \\
\langle\tilde N\rangle_{\rm p}           & 6.673   & 5.939   & 6.093   & 5.918   \\
\langle n^{\rm gm}\rangle_{\rm m}        & 1.525   & 2.447   & 3.744   & 3.885   \\
\langle S^{\rm gm}\rangle_{\rm m}        & 2.579   & 3.534   & 3.573   & 3.357   \\
\hline\hline
\end{tabular}
\label{table:landscape}
\end{center}
\end{table}

%
%
\begin{table}[h]
\caption{The distribution of the number of steps $n^{\rm gm}$ lying
on the shortest path from local minima to the global minimum at
four values of the range parameter $\rho$.}
\begin{center}
\begin{tabular}{Crrrr}
\hline\hline
n^{\rm gm} & \multicolumn{4}{c}{Number of minima} \\
\cline{2-5}
      & $\rho=4$ & $\rho=6$ & $\rho=10$ & $\rho=14$ \\
\hline
1 & 87 & 188 &   71 &  148 \\
2 & 59 & 591 &  937 & 1116 \\
3 & 12 & 518 & 2887 & 3502 \\
4 &    & 116 & 3315 & 4393 \\
5 &    &  19 & 1644 & 2627 \\
6 &    &   6 &  403 &  843 \\
7 &    &     &   47 &  120 \\
8 &    &     &    1 &   10 \\
\hline\hline
\end{tabular}
\label{table:steps}
\end{center}
\end{table}

\newpage

\clearpage
\newpage

\clearpage
\newpage

\clearpage
\newpage

\end{document}